\documentclass[10pt,conference]{IEEEtran}
\IEEEoverridecommandlockouts

\usepackage{amsmath,amssymb,amsfonts}
\usepackage{algorithmic}
\usepackage{graphicx}
\usepackage{textcomp}
\usepackage{xcolor}
\def\BibTeX{{\rm B\kern-.05em{\sc i\kern-.025em b}\kern-.08em
    T\kern-.1667em\lower.7ex\hbox{E}\kern-.125emX}}

\usepackage{algorithm}
\usepackage{booktabs}
\usepackage{enumitem}
\usepackage{listings}
\usepackage{multirow}
\usepackage{subcaption}
\usepackage{xspace}
\usepackage{hyperref}
\usepackage{tabularx}
\usepackage{makecell}
\usepackage[normalem]{ulem}
\usepackage{threeparttable}

\usepackage[final]{mychanges}

\newcommand{\tool}{PALM\xspace}

\definecolor{codeblue}{rgb}{0.02,0.31,0.68}
\definecolor{codeorange}{rgb}{0.58,0.22,0}
\definecolor{codered}{rgb}{0.81,0.13,0.18}
\definecolor{codepurple}{rgb}{0.4,0.22,0.73}
\begin{document}



\title{PALM: Synergizing Program Analysis and LLMs to Enhance Rust Unit Test Coverage\vspace{-0.28cm}}


\author{
\IEEEauthorblockN{
Bei Chu\IEEEauthorrefmark{1},
Yang Feng\IEEEauthorrefmark{1},
Kui Liu\IEEEauthorrefmark{2},
Hange Shi\IEEEauthorrefmark{1},
Zifan Nan\IEEEauthorrefmark{2},
Zhaoqiang Guo\IEEEauthorrefmark{2} and
Baowen Xu\IEEEauthorrefmark{1}
}
\IEEEauthorblockA{\IEEEauthorrefmark{1}\textit{State Key Laboratory for Novel Software Technology, Nanjing University}, Nanjing, China}
\IEEEauthorblockA{\IEEEauthorrefmark{2}\textit{Software Engineering Application Technology Lab, Huawei}, Hangzhou, China}
\IEEEauthorblockA{
\{beichu, hangeshi, gzq\}@smail.nju.edu.cn,
\{fengyang, bwxu\}@nju.edu.cn,
\{kui.liu, nanzifan\}@huawei.com
}
}

\maketitle

\begin{abstract}
Unit testing is essential for ensuring software reliability and correctness. Classic Search-Based Software Testing (SBST) methods and concolic execution-based approaches for generating unit tests often fail to achieve high coverage due to difficulties in handling complex program units, such as branching conditions and external dependencies. Recent work has increasingly utilized large language models (LLMs) to generate test cases, improving the quality of test generation by providing better context and correcting errors in the model's output. However, these methods rely on fixed prompts, resulting in relatively low compilation success rates and coverage.

This paper presents PALM, an approach that leverages large language models (LLMs) to enhance the generation of high-coverage unit tests. PALM performs program analysis to identify branching conditions within functions, which are then combined into path constraints. These constraints and relevant contextual information are used to construct prompts that guide the LLMs in generating unit tests. We implement the approach and evaluate it in \myrep{15}{10} open-source Rust crates. Experimental results show that within just two or three hours, PALM can significantly improve\mydel{s} test coverage compared to classic methods, with increases in overall project coverage exceeding 50\% in some instances and its generated tests achieving an average coverage of \myrep{72.30}{75.77}\%, comparable to human effort (\myrep{70.94}{71.30}\%), highlighting the potential of LLMs in automated test generation. We submitted 91 PALM-generated unit tests targeting new code. Of these submissions, 80 were accepted, 5 were rejected, and 6 remain pending review. The results demonstrate the effectiveness of integrating program analysis with AI and open new avenues for future research in automated software testing.
\end{abstract}


\section{Introduction}

Software testing is crucial in the software development lifecycle, ensuring functionality and compliance with requirements~\cite{planning2002economic, myers2011art, siddiqui2021learning}. Unit testing~\cite{beck2022test} is particularly effective in detecting defects early and maintaining code quality.
Meanwhile, creating unit tests is often time-consuming, typically consuming over 15\% of development time~\cite{daka2014survey, runeson2006survey}. Manually writing test cases is tedious and prone to errors, usually resulting in incomplete coverage and undetected bugs. This underscores the need for automated solutions to generate unit tests efficiently and accurately.
Code coverage is the widely-used metric for evaluating the effectiveness of such automated tests.
It measures the proportion of the source code (e.g., statements, branches) executed by the test suite, providing an objective indicator of its comprehensiveness. Achieving high code coverage is a primary goal in automated test generation, as it increases confidence in the software's correctness by exercising more of its behavior~\cite{kochhar2015code}.

However, classic automated unit test generation methods aimed at achieving high coverage, such as search-based software testing (SBST)~\cite{tymofyeyev2022search, fraser2011evosuite, pacheco2007randoop, lukasczyk2022pynguin, herlim2022citrus} and concolic execution~\cite{takashima2021syrust, garg2013feedback, yoshida2017klover, chen2014test}, have inherent limitations. SBST methods employ search algorithms to generate test inputs or call sequences, retaining those that execute correctly and enhance coverage and then mutate these inputs to maximize coverage.
Despite automating test generation through search optimization, SBST suffers from high computational costs due to repeated program executions~\cite{fraser2011evosuite}, a critical dependency on the fitness landscape which may prioritize coverage over fault detection~\cite{arcuri2010does}, sensitivity to the curse of dimensionality in large input/state spaces~\cite{mcminn2004search}, and difficulties in handling complex program semantics and external dependencies~\cite{machiry2013dynodroid}, limiting its practical efficacy.
Meanwhile, concolic execution techniques, such as Syrust~\cite{takashima2021syrust} and AUTS~\cite{nguyen2024automated}, can satisfy condition constraints and improve coverage by combining concrete and symbolic execution on test drivers. 
Nevertheless, they may suffer from path explosion~\cite{boonstoppel2008rwset, krishnamoorthy2010tackling, kolchin2018novel}, the computational complexity and scalability limitations of constraint solving~\cite{de2008z3}, significant difficulties in accurately modeling environment interactions and external dependencies~\cite{baldoni2018survey, cadar2013symbolic}, and the substantial engineering overhead required for developing and maintaining complex symbolic execution engines~\cite{chipounov2011s2e}.

Recently, large language models (LLMs) have emerged as a promising solution to the limitations of classic test generation methods~\cite{guilherme2023initial, tufano2020unit, deljouyi2024understandable}. LLMs like OpenAI's GPT series~\cite{achiam2023gpt} and DeepSeek~\cite{liu2024deepseek} have demonstrated remarkable capabilities in understanding and generating natural language~\cite{sanh2021multitask, thoppilan2022lamda}. Given that these LLMs are trained on enormous datasets, they can also understand and generate code~\cite{le2022coderl, chen2021evaluating}.
However, existing approaches leveraging LLMs for unit test generation often rely on fixed prompting strategies~\cite{alagarsamy2024a3test, chen2024chatunitest}, lacking explicit guidance on a method's internal structure and feasible execution paths. 
Consequently, when generating tests for methods with complex branch conditions, these approaches often fail to cover longer, intricate paths\textendash a shortcoming that critically undermines the goal of achieving high coverage, a critical testing objective.
Addressing this limitation necessitates techniques that explicitly decompose the problem based on the program's execution paths, enabling targeted guidance to the LLM for each distinct scenario.

Different from existing works, in this paper, we develop PALM, an approach that combines classic program analysis techniques with emerging large language models to generate unit tests satisfying specific path constraints.
Recent research shows that breaking a problem into smaller steps and prompting LLMs to solve each step sequentially yields better results than directly asking for an answer~\cite{wei2022chain, zelikman2022star}. This enables LLMs to tackle more complex problems effectively. Based on this principle, we can decompose the task of generating unit tests into several smaller tasks, each corresponding to a specific execution path within the focal method. This decomposition allows LLMs to focus on specific subtasks, resulting in more accurate and diverse test cases. By combining these tests, we can create a comprehensive test suite that effectively covers various execution paths.

PALM begins with path analysis of the focal method to identify its execution paths, each representing a sequence of conditions and return values. We gather contextual information, including the focal method's body, other declarations, definitions, and function call details, to guide the LLM. For each path, we construct a prompt using this information and ask the LLM to generate test cases. This ensures focused, comprehensive test case generation. We have implemented PALM for Rust and evaluated it on \myrep{15}{10} open-source projects with \myrep{4145}{3219} focal methods. The GPT-4o mini-based prototype achieved an increase of over 20\% in the generation of compilable tests and an improvement in coverage exceeding 50\%. 91 unit tests, generated by PALM to cover code regions lacking prior developer-written tests, were submitted via pull requests. Subsequently, 80 of these tests were integrated into the project, representing an 94.12\% acceptance rate among those reviewed.

In summary, this paper makes the following contributions:

\begin{itemize}[leftmargin=0.6cm]

\item \textbf{Approach.} We introduce PALM, which combines program analysis with LLMs to generate high-coverage unit tests. This approach breaks the task into steps based on specific conditional paths within the focal method, using targeted prompts to guide the LLM.

\item \textbf{Tool.} We implemented the PALM based on rustc APIs and designed a system that automatically generates unit tests through interaction with a large language model. The system is currently a prototype, with plans for open-sourcing upon completion.

\item \textbf{Study.} We evaluated PALM on \myrep{15}{10} open-source projects with \myrep{4145}{3219} focal methods. Our GPT-4o mini-based prototype increased the ratio of correctly generated, compilable tests by up to 20\% and improved coverage by up to 50\%.

\item \textbf{Practical Usage.} 91 generated unit tests were submitted as pull requests to 8 distinct, popular open-source Rust projects. Project maintainers subsequently accepted and merged 80 of these submissions, accompanied by positive feedback. This represents an acceptance rate of 94.12\% among the unit tests that underwent maintainer review.

\end{itemize}


\section{Background}

\subsection{Unit Test Generation}

Unit test generation automates test creation for methods under test (focal methods), addressing the time-consuming and error-prone nature of manual testing~\cite{daka2014survey, runeson2006survey}. Classic approaches fall into two categories: search-based~\cite{herlim2022citrus, fraser2013evosuite, tymofyeyev2022search, lukasczyk2022pynguin} and symbolic execution-based~\cite{takashima2021syrust, yoshida2017klover, garg2013feedback, rho2024taming}.

\textbf{Search-Based Software Testing (SBST).} 
Search-Based Software Testing (SBST) methods uses meta-heuristic algorithms (e.g., genetic algorithms) to iteratively mutate test cases for coverage improvement~\cite{mcminn2004search, mcminn2011search}. Fundamentally, these tools operate by leveraging two core steps: (i) initially generating a random set of test cases as seed inputs and (ii) iteratively mutating these test cases to enhance the coverage of the focal method. Thanks to search algorithms, SBST methods can automatically explore possible test inputs and generate tests with higher coverage through optimized search techniques.
However, SBST methods face the following challenges: (i) when dealing with complex and large systems, the search algorithms can consume substantial computational resources~\cite{harman2008search}, (ii) they can get trapped in local optima~\cite{fraser2012whole}, and (iii) the quality of the generated tests is influenced by the initial inputs and fitness landscape characteristics~\cite{harman2015achievements}.

PALM decomposes the test generation task based on identified execution paths, each associated with specific branch condition constraints. This decomposition significantly reduces the search space that needs to be considered for each sub-task, thereby reducing computational resource consumption compared to SBST. Additionally, by leveraging LLMs' ability to understand code semantics and constraints, PALM can generate tests that are more directly aligned with the requirements of a specific path, reducing the reliance on random initial inputs and their potential impact on result quality.

\textbf{Concolic Execution.}
Concolic execution blends concrete and symbolic execution to collect path constraints and generate tests~\cite{baldoni2018survey, sen2005cute, godefroid2005dart}. It uses a constraint solver to reason about path constraints, thereby generating test inputs. Through symbolic execution, this approach can systematically explore different execution paths in a program, generating high-coverage test cases. However, this method still faces several challenges in practice. Loops in the program can lead to path explosion, causing symbolic execution to consume substantial computational resources. Constraint solvers may also struggle with complex constraint conditions. Furthermore, symbolic execution often finds handling code involving external dependencies, such as file systems or networks challenging.

Similar to symbolic execution, PALM leverages path analysis to identify the constraints defining each execution path. However, instead of feeding these constraints to a classic solver, PALM uses the constraints as part of a natural language prompt to guide an LLM to generate concrete input values that satisfy the constraints for that specific path. This approach aims to bypass the computational complexity of constraint solving and the challenges in modeling external dependencies inherent in classic symbolic execution engines.

\subsection{LLM for Code Generation}

Large language models (LLMs) have demonstrated impressive capabilities in the field of natural language processing (NLP), excelling in various tasks such as conversation~\cite{thoppilan2022lamda}, text generation~\cite{brown2020language, chowdhery2023palm, shoeybi2019megatron}, and reasoning~\cite{cobbe2021training}. Given their success in NLP, attention has naturally turned to code-related tasks, leading to the creation of numerous code generation models, such as AlphaCode~\cite{li2022competition}, PolyCoder~\cite{xu2022systematic}, Codex~\cite{chen2021evaluating}, Google's program synthesis model~\cite{austin2021program}, and CodeGen~\cite{nijkamp2022conversational}.

The input text provided to the LLM during inference is called a prompt, which guides the model's generation process. Through prompt engineering, researchers can accomplish various code-related tasks without the need to retrain the model, such as code completion~\cite{ziegler2022productivity}, code explanation~\cite{sarsa2022automatic}, and code repair~\cite{pearce2021can, prenner2021automatic, deligiannis2023fixing}. In this work, we apply prompt engineering techniques to construct structured prompts that explicitly incorporate the \mydel{extracted }path constraints and context for each execution path, guiding the LLM to generate test cases.

\section{Definitions and Motivating Example}
\label{sec:example}

We present the definitions used throughout the paper and illustrate our objectives with an example.

\textbf{Context.} In this paper, the term \textit{context} generally refers to the code related to, but distinct from, the focal method and the test functions. There are two types of context: (i) \textit{focal context}, which is the context of the focal method, typically consisting of headers, macros, and global variables in the file containing the focal method; the declarations and definitions of the class to which the focal method belongs (if applicable); and the declarations and definitions of the data types and functions used by the focal method. (ii) \textit{test context}, which is the context of the test functions, usually including the necessary use statements for the test files and any predefined data structures and variables required for the tests.

\textbf{Condition Chain.} By recording the branch conditions that must be satisfied for control flow transitions during path analysis, we can obtain approximate path constraints, which are sequences of conditions that must be met for a given execution path. In this paper, we refer to such sequences of conditions as condition chains.

\begin{lstlisting}[caption=Motivating examples for \texttt{wait} function. Details of struct/trait definitions are omitted., label=code:wait]
fn wait(queue, mut curr_queue) {
    loop {
        if let Err(new_queue) = exchange {
            if strict::addr(new_queue) & STATE_MASK != curr_state {
                return;
            }
            continue;
        }
        while !node.signaled.load(Ordering::Acquire) {
            thread::park();
        }
        break;
    }
}
\end{lstlisting}

\begin{figure}[tbp]
    \centering
    \includegraphics[width=0.9\linewidth]{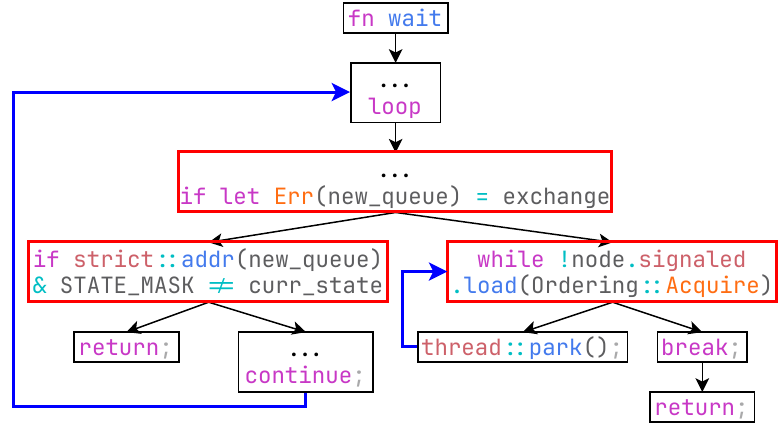}
    \caption{Control flow graph of the focal method \texttt{wait}\myrep{, highlighting the complex control flow involving conditional branches and nested loops that pose significant challenges for automated test generation}{ with highlighted conditional branches and nested loops}.}
    \label{fig:cfg}
\vspace{-0.5cm}
\end{figure}

Next, we illustrate the challenges current test case generation methods face using the following example. Code~\ref{code:wait} presents the \texttt{wait} method from the \texttt{once\_cell} crate~\cite{githubGitHubMatkladonce_cell}, which is used to block the current thread until a concurrent initialization completes\mydel{, employing a lock-free queue of waiters to coordinate thread synchronization}. To better depict the method's control flow behavior, we present its control flow graph (CFG) in Fig.~\ref{fig:cfg}. The CFG of the \texttt{wait} method reveals a structured composition of 2 conditional branches and 2 nested loops\mydel{, orchestrating thread synchronization through lock-free operations}. To ensure comprehensive coverage, test inputs must be carefully crafted to meet the constraints.

\myadd{Despite its power, symbolic execution struggles with functions like \texttt{wait}. Its nested loops can trigger a path explosion, overwhelming the analysis~\cite{cadar2013symbolic}, while Rust's unique ownership model and concurrency primitives like AtomicPtr pose significant modeling challenges for symbolic engines, hindering scalability~\cite{anand2013orchestrated, bucur2015improving}.
Meanwhile, existing LLM-based approaches often fail because they (i) lack the specific context to correctly set up the function's preconditions~\cite{chen2021evaluating}, (ii) cannot generate the precise inputs needed to navigate the intricate control flow for high coverage~\cite{allamanis2018survey}, and (iii) frequently produce code that violates Rust's strict ownership and type safety rules, leading to low compilation rates. PALM is designed to overcome these specific challenges by synergizing program analysis with LLM generation.}

\begin{figure*}[tbp]
    \centering
    \includegraphics[width=\linewidth]{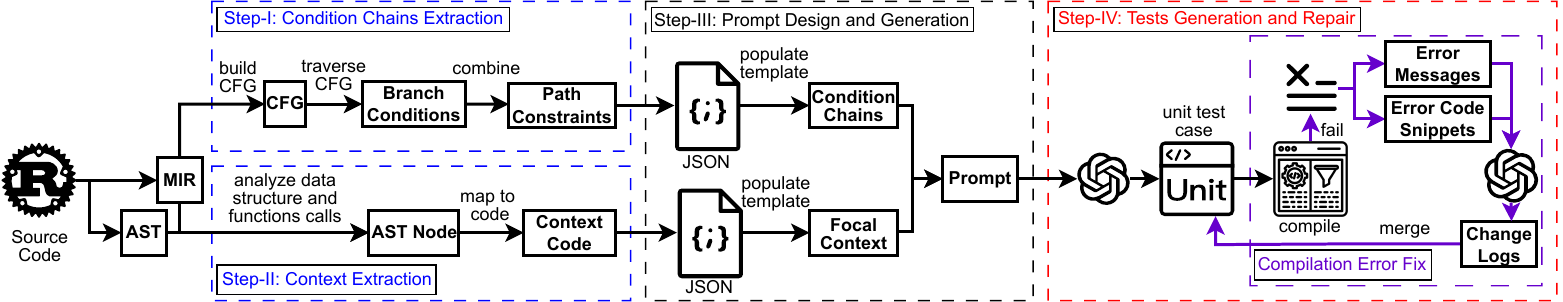}
    \caption{\myrep{The PALM framework. Step I extracts path constraints from the MIR-level CFG. Step II gathers contextual code via AST/IR analysis. Step III synthesizes a path-specific prompt. Step IV generates tests and uses iterative, compiler-guided repair}{Overview of the automated test generation framework: extracting (I) path constraints and (II) method contexts, (III) synthesizing LLM prompts, then (IV) generating executable tests via input range deduction, test prefix construction, and oracle generation with iterative compiler feedback}.}
    \label{fig:framework}
\vspace{-0.5cm}
\end{figure*}

\mydel{Despite its power in automated test generation, symbolic execution struggles with complex functions like \texttt{wait} due to several inherent challenges: (i) \textit{path explosion}: Loops within the function can cause a combinatorial explosion of execution paths. As each iteration can create a new path, the total number of paths may grow exponentially, quickly becoming unmanageable~\cite{cadar2013symbolic}. (ii) \textit{ownership and borrowing}: Rust's unique ownership model and borrow checker pose significant hurdles. Symbolic execution must correctly handle smart pointers (like Box, Rc) and track ownership transfers to ensure memory safety, a non-trivial task during test generation~\cite{anand2013orchestrated}. (iii) \textit{scalability}: Consequently, path explosion and the overhead of managing complex data structures render symbolic execution computationally intensive. This severely curtails its applicability to larger and more complex codebases~\cite{bucur2015improving, baldoni2018survey}.}

\mydel{Meanwhile, existing methods leveraging LLMs for test generation also encounter several challenges when applied to \texttt{wait}:
(i) \textit{lack of context.} LLMs often generate test cases without comprehensively understanding the code's context. The \texttt{wait} method interacts with fields of the \texttt{Waiter} struct and invokes auxiliary methods like \texttt{strict::addr}, requiring contextual awareness to generate semantically meaningful test cases. Without this context, LLMs may fail to adequately exercise the function's critical paths~\cite{chen2021evaluating}.
(ii) \textit{low coverage.} Achieving high code coverage is crucial for effective testing. Yet, LLMs may struggle to generate inputs that cover all execution paths, particularly in the presence of loops and complex conditional logic. The \texttt{wait} method's nested loop structure for atomic queue manipulation and its conditional branches for handling contention and state validation pose significant challenges for achieving full path coverage~\cite{allamanis2018survey}.
(iii) \textit{low compilation success rate.} Test cases generated by LLMs may exhibit syntactic or semantic inconsistencies, resulting in low compilation success rates. This challenge is especially evident in Rust methods like \texttt{wait}, where the strict enforcement of ownership rules, borrowing constraints, and type safety requires meticulously crafted and error-free inputs.}

\mydel{These challenges underscore the limitations of both symbolic execution and LLM-based approaches in generating comprehensive test cases for functions with intricate control flows and data structures. We aim to build upon existing LLM-based test generation methods by incorporating static analysis to extract additional information from focal methods. By leveraging this information, we strive to enhance the capabilities of LLMs, thereby improving the compilation success rate and coverage of the generated test cases.}

\section{Methodology}

In this section, we demonstrate the methodology for generating unit tests for a focal method, such as \texttt{wait}. PALM involves generating prompts corresponding to the execution paths within the focal method. This breaks down the task of generating a comprehensive test suite into several steps, each focused on generating tests for a single path. These prompts ultimately guide the model in generating unit tests with high coverage.
When test cases fail to compile, we provide the model with erroneous code snippets and error messages, enabling it to suggest repairs iteratively\mydel{ and improve the compilation success rate}.

\subsection{Overview}

The core of PALM involves collecting the chain of conditions along each path in the focal method through path analysis. By leveraging this technique, we can capture the program's behavior and gather the constraints along the execution paths.

Classic test generation methods based on symbolic execution enumerate all possible path constraints and generate test cases by solving these constraints~\cite{godefroid2005dart}. However, real-world code often includes complex data structures and external dependencies, making constraint-solving challenging. Loops or deeply nested conditions can also lead to path explosion, consuming significant computational resources.

PALM addresses these challenges threefold. First, it retains complex constraints during condition chain collection, avoiding extensive solving. Second, it leverages the LLM's understanding of constraints and context for test case generation, instead of relying on constraint solvers. Finally, to reduce computational overhead, PALM selects a minimal path set covering all branch conditions, tackling this NP-hard set cover problem with a greedy approximation algorithm~\cite{alon2003online}.
\myadd{The number of generated tests is determined by the control flow complexity of the focal method. By default, PALM generates one test per minimized path. Additionally, it may generate one or two boundary tests for path conditions, budget permitting. The decision is based on the nature of the conditional expression: one test for single-boundary conditions (e.g., \texttt{x > 10}) and two for range-based conditions (e.g., \texttt{x >= 0 \&\& x <= 10}). This strategy is designed to achieve full branch coverage efficiently while also probing critical edge cases.}

Therefore, our test generation process consists of four steps: (I) collecting condition chains of the focal method; (II) gathering and constructing context for the focal method; (III) designing and populating a prompt template suitable for an LLM with the condition chains and context; and (IV) using the prompt to guide the LLM in generating unit tests and iteratively fixing any test cases that fail to compile. Fig.~\ref{fig:framework} provides an overview of PALM to test generation.

\subsection{Step-I: Condition Chains Collection}
\label{sec:chaincoll}

We aim to generate tests that satisfy specific execution paths within the focal method. To achieve this, we must collect the relevant condition chains and provide this information to the model. This step involves a detailed explanation of how we traverse the focal method's Control Flow Graph (CFG) to gather the necessary condition chains.

\subsubsection{Path Constraints Collection}

Our approach leverages both the MIR and the HIR to derive conditional constraints from Rust source code. Unlike existing tools like MirChecker~\cite{li2021mirchecker}, which focus on flow-sensitive security analyses, our method is specifically designed to recover high-level semantics for unit test generation. MIR, a simplified version of Rust code, is typically used for flow-sensitive security analyses. It is constructed based on control flow graphs (CFGs), where nodes represent basic blocks—sequences of instructions without branches or jumps—and edges represent control flow transitions, capturing potential execution paths and their associated conditional constraints. By analyzing these control flow transitions, we can extract the underlying conditional constraints. HIR is a compiler-friendly abstraction of the Abstract Syntax Tree (AST) and is closer to the syntactic structure of Rust source code compared to MIR.

Using the rustc API, we obtain the MIR and HIR of the focal method and perform a preorder traversal of the CFG to explore all possible execution paths. When encountering control flow transitions, we analyze the terminator of the basic block to extract and record the conditional constraints. Upon reaching the exit basic block, we record the return value, thereby constructing a conditional chain for the execution path.

\begin{algorithm}
\caption{Path Minimization}
\label{alg:path_mini}
\begin{algorithmic}
\REQUIRE set of path constraints $P$
\ENSURE set of minimized path constraints $M$
\STATE $M \gets \emptyset$
\STATE $S \gets $ getConstraints($P$)
\WHILE{$S \neq \emptyset$}
    \STATE $p_m \gets \underset{p \in P}{\arg\max} \, |\text{getConstraints}(p) \cap S|$
    \STATE $M \gets M \cup \{p_m\}$
    \STATE $S \gets S \setminus $getConstraints($p_m$)
\ENDWHILE
\end{algorithmic}
\end{algorithm}

\subsubsection{Path Minimization}
\label{sec:path_mini}

\begin{table*}[tbp]
    \caption{Condition Chains of \texttt{wait}}
    \label{tab:condition_chains}
    \centering
    \newcommand{\red}[1]{\color{red}{#1}}
    \resizebox{\linewidth}{!}{
    \begin{tabular}{cccccc}
        \toprule
        \multirow{2}{*}{Path} & \multicolumn{5}{c}{Conditions} \\
        \cmidrule{2-6}
        & \makecell[l]{let Err(new\_queue) \\ = exchange}
        & \makecell[l]{strict::addr(new\_queue) \\ \& STATE\_MASK != curr\_state}
        & \makecell[l]{let Err(new\_queue) \\ = exchange$^{\mathrm{*}}$}
        & \makecell[l]{node.signaled.load \\ (Ordering::Acquire)}
        & \makecell[l]{node.signaled.load \\ (Ordering::Acquire)$^{\mathrm{*}}$} \\
        \midrule
        \red{0} & false & \textemdash & \textemdash & true & \textemdash \\
        1 & false & \textemdash & \textemdash & false & true \\
        \red{2} & true & true & \textemdash & \textemdash & \textemdash \\
        3 & true & false & false & true & \textemdash \\
        \red{4} & true & false & false & false & true \\
        \bottomrule
        \multicolumn{6}{l}{$^{\mathrm{*}}$Loop-Terminating constraints.}
    \end{tabular}
    }
\vspace{-0.5cm}
\end{table*}


The number of possible execution paths in a method grows exponentially with the number of branches, making it impractical to generate test cases for all paths. To address this challenge, we aim to identify a minimal set of paths that can cover all branch conditions. Our approach decomposes the problem into two stages: path traversal and set cover. During path traversal, we collect all possible paths without filtering to ensure completeness. Subsequently, we formulate the problem of selecting a minimal set of paths as a classic minimum set cover problem.

Algorithm~\ref{alg:path_mini} details our approximate solution. It inputs a set of path constraints, each comprising multiple condition constraints, and outputs a minimal subset of these paths covering all condition constraints.
The algorithm greedily constructs the minimized set $M$. It begins with a set $S$ containing all unique branch conditions to be covered. In each iteration, it selects the path from the original set $P$ that covers the maximum number of conditions currently in $S$. This path is added to $M$, and its corresponding conditions are removed from $S$. The process repeats until $S$ is empty.
By solving this problem, \tool significantly reduces the number of paths that need testing while ensuring complete branch coverage.

For example, consider the function \texttt{wait} shown in Code~\ref{code:wait}, after performing path analysis, we obtained the five condition chains illustrated in Table~\ref{tab:condition_chains}.
For the five paths in \texttt{wait}, we first select the path that covers the most conditions, path 4. Next, we choose path 0, which covers most of the remaining conditions, and finally, path 2. Thus, after minimization, we have selected three of the five paths marked red in Table~\ref{tab:condition_chains}.

\subsection{Step-II: Context Construction}

Previous work on generating tests using LLMs has shown that incorporating additional context in the prompt enhances the quality of the generated tests~\cite{tufano2020unit}. To leverage this insight, we extract and incorporate relevant context through analysis. This context is categorized into two types:

\begin{itemize}[leftmargin=0.6cm]
    \item \textit{Structure Context}: Definitions or declarations of member variables, methods, and structure fields associated with the focal method. This helps the model understand the focal method's role within its class or structure.
    \item \textit{Dependency Context}: Declarations or definitions of parameter types, data types, and functions directly or indirectly used by the focal method. This ensures the generated code correctly handles external dependencies.
\end{itemize}

To manage token limits and reduce redundancy, we provide full definitions for directly used types and methods, using only relevant declarations for indirect dependencies.

The context extraction begins by parsing the focal file to extract use statements, global definitions, and type/function definitions, with use statements and global definitions being added to the context. If the focal method resides within a struct, the struct's definition and implementations (fields, associated functions, directly called methods) are extracted, while other methods receive only signatures. Definitions of direct dependencies\textendash types and functions directly used by the focal method\textendash are added. For indirect dependencies, local variable types and function calls are extracted by analyzing each function/method's Intermediate Representation (IR). Subsequently, a crate-wide Abstract Syntax Tree (AST) parse resolves use and mod dependencies, yielding a structured representation including trait bounds and return types. The final context aggregates file-level use/globals, direct/indirect dependencies (including struct/enum definitions, trait implementations), and information from call-graph/crate-wide analyses (e.g., trait bounds, return types). This assembled AST is then converted to Rust source code suitable for the LLM prompt.

\subsection{Step-III: Prompts Design and Construction}

\begin{figure}[tbp]
    \centering
    \includegraphics[width=\linewidth]{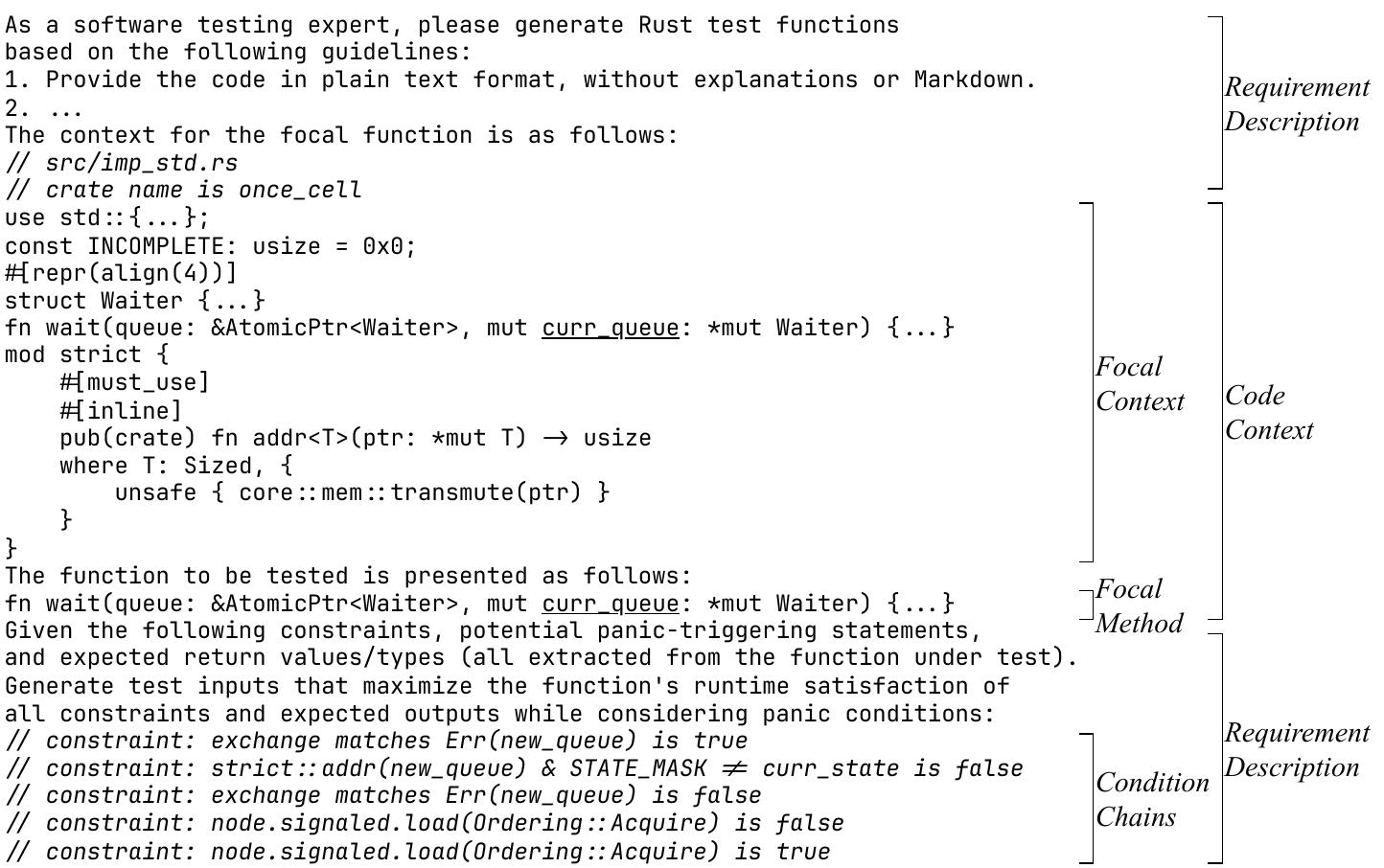}
    \caption{\myrep{Structure of a PALM-generated prompt for the \texttt{wait} function. The Requirement Description part instructs the LLM and provides the target condition chain. The Code Context part supplies source code along with curated contextual definitions}{Prompt structure for testing \texttt{wait}: requirement description part (role specification, task instructions, condition chains) and code context part (focal context, focal method). In this case, no additional test context is required beyond the method's immediate scope}.}
    \label{fig:prompt}
\vspace{-0.5cm}
\end{figure}

Once the condition chains and context for the focal method are collected, we construct the prompt based on the focal method body, path constraints, return values, and other contextual information. To balance between overly simplistic prompts that might underestimate LLMs’ capabilities and overly complex prompts that are impractical, we design our prompt by following established practices in unit test generation research~\cite{yuan2024evaluating, chen2024chatunitest} and widely-recognized experiences in using ChatGPT~\cite{ChatGPTG96:online}. Our prompt consists of two main components: (i) the requirement description part, which explains the task to LLMs, and (ii) the code context part, which includes the focal method and relevant context. An example for \texttt{wait} is shown in Fig.~\ref{fig:prompt}.

\textbf{Code Context.} Inspired by previous work~\cite{tufano2020unit, yuan2024evaluating} on generating tests using LLMs and leveraging the content we have collected, the code context part includes the following components: (i) \textit{focal context.} This is the context related to the focal method. (ii) \textit{focal method.} This includes the signature and the body. (iii) \textit{test context.} In some cases, the test context must also be included, as it comprises the necessary use statements and type declarations required for test execution.

\textbf{Requirement Description.} Based on established practices for using ChatGPT~\cite{yuan2024evaluating}, our requirement description part includes (i) \textit{role-playing instruction}, e.g., ``as a software testing expert,'' to prime GPT for test generation~\cite{ChatGPTG96:online, dong2023self}. (ii) \textit{task description instructions}, such as ``please generate test functions based on the following guidelines:''. To ensure the chat-tuned model generates only the required test cases without other descriptions, we include instructions like ``provide the code in plain text format'', and ``without explanations.'' (iii) \textit{condition chains}, to ensure that the model generates appropriate test cases. By structuring the prompt this way, we aim to maximize the model's ability to generate high-quality unit tests.

\subsection{Step-IV: Test Generation and Repair}
\label{sec:gen_repair}

\begin{figure}[tbp]
    \centering
    \includegraphics[width=\linewidth]{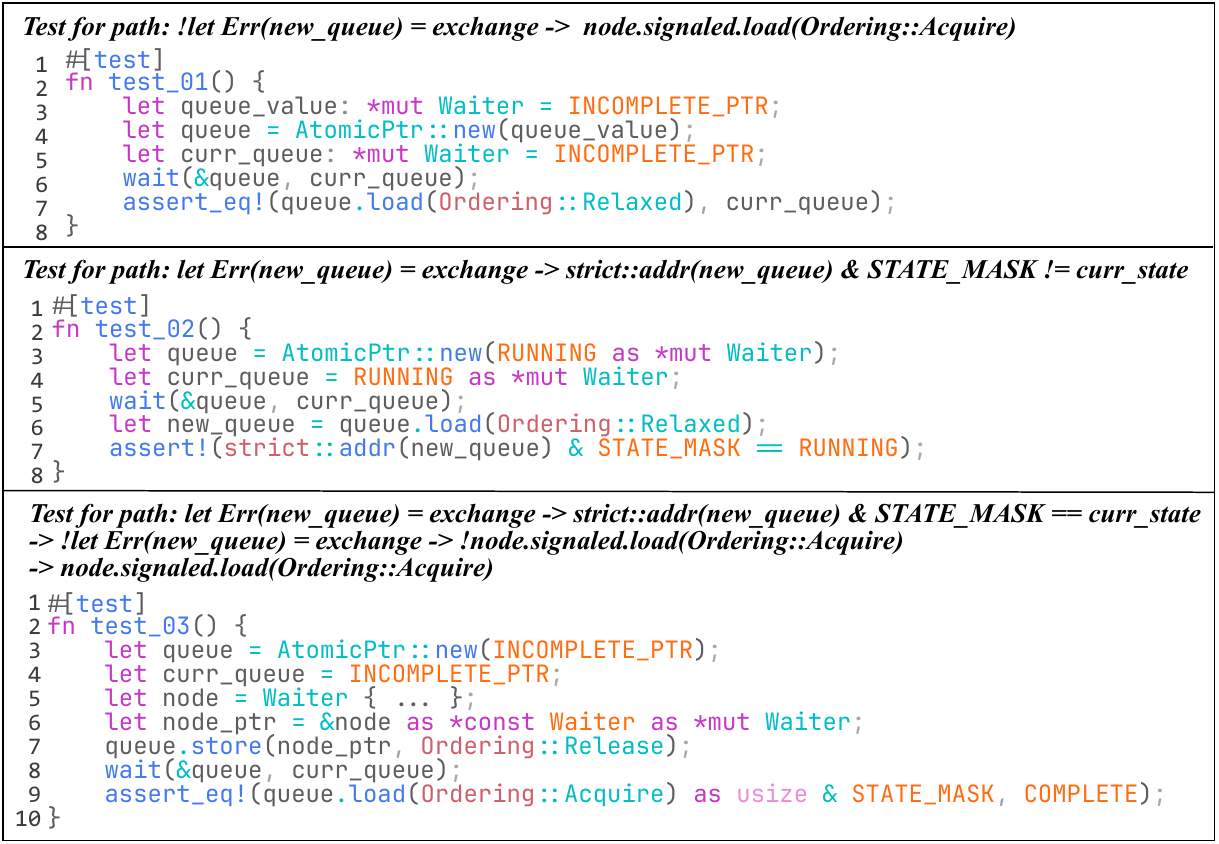}
    \caption{\myrep{Example tests for \texttt{wait} generated by PALM, with one test shown per targeted execution path. Each test is crafted to satisfy a distinct sequence of predicates}{Example tests for \texttt{wait} generated from GPT-4o mini, with one test selected per path}.}
    \label{fig:test_cases}
\vspace{-0.5cm}
\end{figure}

After constructing the initial prompt, we employ an iterative process to guide the model in generating test cases for each execution path. This decomposition by individual paths facilitates a comprehensive, high-coverage test suite.
Fig.~\ref{fig:test_cases} illustrates the  test cases generated for \texttt{wait} using GPT-4o mini, demonstrating the effectiveness of our approach.

A significant challenge in generating test cases with LLMs is that the generated tests may fail to compile, compromising their effectiveness and coverage. Although truncating uncompilable tests generated by LLMs until they achieve successful compilation~\cite{ryan2024code} might appear as a plausible solution, this practice often results in tests that are incomplete or even semantically invalid, consequently leading to diminished code coverage. A more robust solution is leveraging the model to fix compilation errors. Two established methods are: (i) providing the error information and context to the model, allowing it to regenerate the tests until they compile or a limit is reached~\cite{yuan2024evaluating}; and (ii) providing the error information and code snippets to the model while requesting a change log, which is then used to replace the erroneous code~\cite{deligiannis2023fixing}.

PALM adopts the second method for fixing compilation errors, as regenerating tests may inadvertently alter error-free parts. First, we extract error information and relevant code snippets from the compilation messages. For each error, we construct a prompt instructing the model to generate a change log based on this information. The code replacements are extracted from the model's response and applied before recompiling. This process iterates until the error is resolved or the iteration limit $I$ is reached. To prevent the repair process from becoming excessively time-consuming, we also set an upper limit $E$ on the number of errors to be fixed for each test function. In our experiments, we set $E$ and $I$ to 10 and 3, respectively. Our prompt design is inspired by prior work on using LLMs to fix Rust compilation errors~\cite{deligiannis2023fixing}, but we have streamlined the repair process to improve efficiency.

\begin{figure}[tbp]
    \centering
    \includegraphics[width=\linewidth]{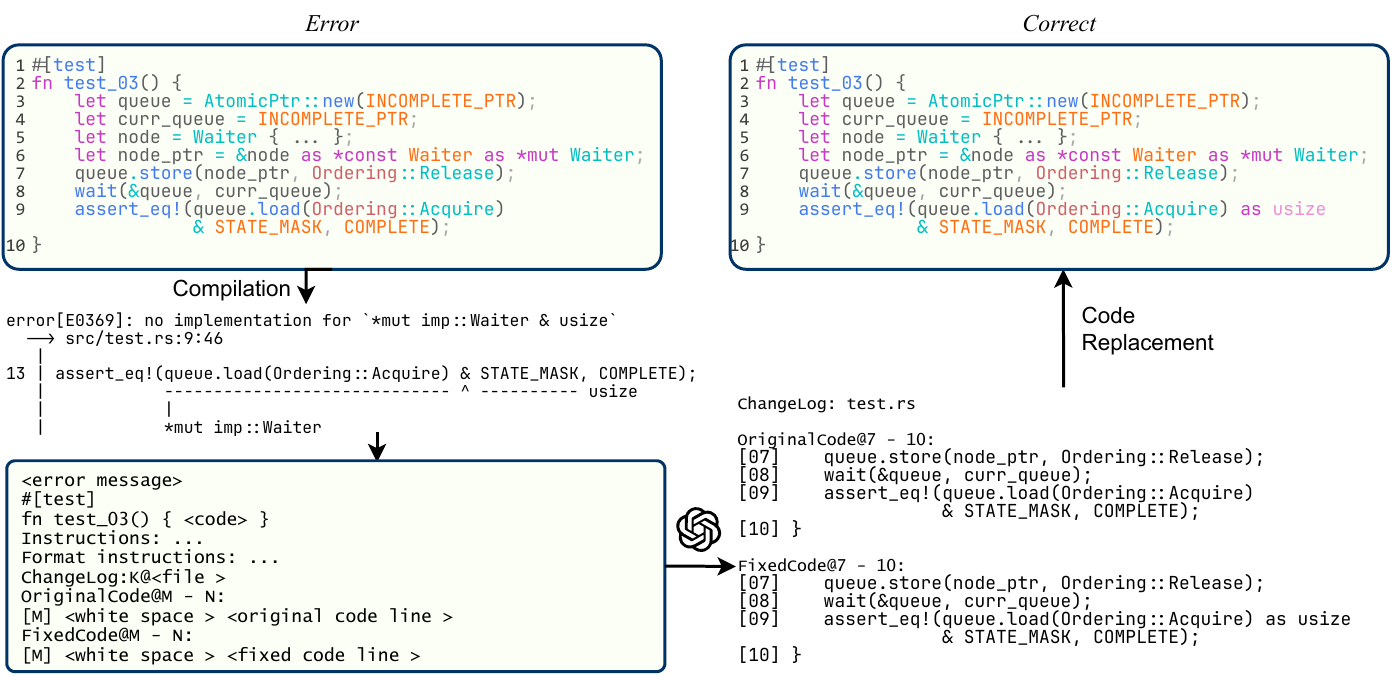}
    \caption{\myrep{The iterative repair process. 1) Capture the error message and surrounding code. 2) Prompt the LLM to generate a change log for a minimal fix. 3) The change is applied, and the code is recompiled. This loop continues until compilation succeeds or a predefined budget is exhausted}{Compilation error fixing process with LLM for \texttt{wait}: given error messages and code snippets, the LLM generates change logs which are iteratively applied until all errors are resolved or the maximum iteration limit is reached}.}
    \label{fig:fix}
\vspace{-0.5cm}
\end{figure}

Fig.~\ref{fig:fix} illustrates fixing a compilation error in a test generated for the \texttt{wait} function. Initially, we compile the test and encounter a compilation error. We extract the error's location, line number, and message from the compiler output. In this example, the error occurs at line 4, with the message: error[E0369]: no implementation for \texttt{`}*mut imp::Waiter \& usize\texttt{`}. We then populate a prepared template with the error information and several lines of code surrounding the error. The model returns a corresponding change log based on the provided prompt format. From this change log, we identify the necessary code modifications. The correction involves adding the type conversion \texttt{as usize}, resolving the issue.

\section{Experiment Design, Results and Discussion}

We address the following RQs in our evaluation:

\begin{itemize}
    \item \textbf{RQ 1. Effectiveness.} How effective is PALM in generating high-quality unit tests compared with baselines?
    \item \textbf{RQ 2. Ablation Study.} What is the impact of condition chains, context, and iterative compilation error repair on the effectiveness of generated unit tests?
    \item \textbf{RQ 3. Practical Usability.} How is PALM's usability for real world applications?
\end{itemize}

\textbf{Experiment Settings.} All our evaluations were conducted on a system equipped with an Intel Core i7-13700k processor, 64GB of RAM, running Ubuntu 24.04. The implementation of the path analysis and context extraction components of the PALM approach is based on Rust version nightly-2024-07-21. We validated PALM on GPT-4o mini~\cite{achiam2023gpt}, an advanced language model developed by OpenAI, designed for faster and more efficient natural language processing tasks.

\textbf{Evaluation Metrics.} Based on previous work in test generation~\cite{yuan2024evaluating, tufano2020unit, ryan2024code}, we have selected the following metrics:

\begin{enumerate}[leftmargin=0.6cm]
    \item \textbf{\# Tests:} The number of generated tests for a project.
    \item \textbf{Comp Pass:} The proportion of tests that can compile.
    \item \textbf{Exec Pass:} The proportion of tests that successfully execute and pass.
    \item \textbf{Line \& Branch Coverage:}
    This metric reflects the effectiveness of the whole generated test suites. Line coverage is defined as the ratio of the number of lines of code executed to the total number of lines of code in the project after running all generated tests. Similarly, branch coverage is the ratio of the number of branches covered to the total number of branches.
\end{enumerate}

\textbf{Subject Programs.}
We selected \myrep{15}{10} open-source Rust crates from crates.io~\cite{cratesio} \myadd{and GitHub} to serve as our subject programs. \myrep{These projects are chosen to form a challenging and representative benchmark, selected for their maturity, high popularity (most with over 300 million downloads), and domain diversity. They cover a wide range of applications, from data serialization and complex data structures to low-level systems programming and core algorithms, aiming to support the generalizability of our findings}{These projects are chosen for their popularity and widespread use in various algorithmic and utility contexts}. In total, we selected \myrep{4145}{3219} focal methods to ensure a robust evaluation. Table~\ref{tab:benchmark} provides detailed information about these \myrep{15}{10} crates, including their download counts, sizes, and other relevant characteristics.

\textbf{Baselines.}
\myadd{We selected five baselines to compare with PALM: two classic tools and three LLM-based approaches. The classic tools are RustyUnit~\cite{tymofyeyev2022search}, a search-based test suite generator for Rust, and RuMono~\cite{zhangrumono}, a fuzzing-based tool that extends RULF~\cite{jiang2021rulf} with capabilities for generic APIs. The LLM-based baselines include: RUG~\cite{rug}, which lacks path-guided prompts and LLM-based repair, relying on fuzzing to enhance coverage; SymPrompt~\cite{ryan2024code}, which differs from PALM by using AST-based path analysis (instead of CFG-based) and truncating failing tests (instead of repairing them); and ChatTester~\cite{yuan2024evaluating}, which contrasts with PALM's path-decomposed generation and patch-based repair by using one-shot generation and full regeneration for fixes.}
\mydel{We selected three baselines to compare with PALM to validate its effectiveness: RustyUnit~\cite{tymofyeyev2022search}, a search-based test suit generator for Rust; RuMono~\cite{zhangrumono}, a fuzzing-based tool for testing Rust APIs that extends RULF~\cite{jiang2021rulf} with enhanced capabilities for generic APIs; and RUG~\cite{rug}, a unit test generation method based on GPT. The differences between RUG and PALM are: (i) RUG does not include condition chains related to path constraints in the prompt, (ii) unlike PALM, RUG does not employ LLMs for test compilation error repair, and (iii) RUG enhances coverage through fuzzing, while PALM systematically generates tests targeting distinct execution paths of focal methods.}

\subsection{RQ1: Effectiveness of PALM for Unit Test Generation}

We evaluated PALM against \myrep{five}{three} baseline tools\mydel{\textendash RustyUnit (SBST), RuMono (fuzzing), and RUG (LLM-based)\textendash} on the subject crates detailed in Table~\ref{tab:benchmark}. For RuMono, we set a 24-hour timeout. We first measured the number of generated test artifacts \myrep{(fuzz targets for RuMono; unit tests for others)}{(unit tests for RustyUnit, RUG, PALM; fuzz targets for RuMono)}. We then assessed their compilation success rates. Successfully compiled tests were subsequently executed to determine their pass rates. Finally, we measured line and branch coverage for each project. Furthermore, we quantified the coverage achieved by the manually-authored unit tests within each selected project. Given the mature, popular, and extensively utilized nature of these projects, their existing test coverage levels were considered indicative of high-quality, manually-derived test suites achievable in practice.




\begin{table}[tbp]
    \caption{Benchmark Crates}
    \label{tab:benchmark}
    \centering
    \scriptsize
    \resizebox{\columnwidth}{!}{
    \begin{tabular}{lllll}
    \toprule
    \textbf{Crate} & \textbf{Description} & \textbf{Downloads} & \textbf{\# Files} & \textbf{LOC} \\
    \midrule
    base64 & Base64 decoding and encoding & 567M & 30 & 5k \\
    bytes & Utility library for working with bytes & 347M & 33 & 6k \\
    fastrand & Random number generator & 302M & 5 & 833 \\
    hashbrown & Rust port of SwissTable hash map & 629M & 35 & 11k \\
    \myadd{http} & \myadd{Library of common HTTP types} & \myadd{304M} & \myadd{34} & \myadd{10k} \\
    indexmap & Pure-Rust hash table & 486M & 31 & 9k \\
    once\_cell & One-time initialization for globals & 439M & 20 & 2k \\
    \myadd{qsv} & \myadd{Blazing-fast Data-Wrangling toolkit} & \myadd{225K} & \myadd{217} & \myadd{85k} \\
    quote & Turn Rust syntax tree into source code & 582M & 19 & 2k \\
    rand & Random generators & 516M & 60 & 9k \\
    regex & Search strings for matches of a regex & 432M & 222 & 97k \\
    \myadd{ryu} & \myadd{Pure Rust implementation of Ryū} & \myadd{418M} & \myadd{29} & \myadd{3k} \\
    serde & Framework for serializing and deserializing & 517M & 188 & 31k \\
    serde\_json & JSON serialization file format & 451M & 70 & 17k \\
    strsim-rs & String similarity metrics & 390M & 3 & 1k \\
    \bottomrule
    \end{tabular}
    }
\vspace{-0.5cm}
\end{table}

\begin{table*}[tbp]
\centering
\caption{\myrep{Assessment of Generated Tests: Volume, Compilation, and Execution Success.}{Evaluation of different Approaches}}
\label{tab:RQ1_a}
\newcommand{\data}{\textemdash}
\newcommand{\tb}[1]{\textbf{#1}}
\newcommand{\tu}[1]{\uline{#1}}
\newcommand{\tn}[1]{\tnote{#1}}
\resizebox{\linewidth}{!}{
\begin{threeparttable}
\begin{tabular}{l*{6}{c}|*{6}{c}|*{4}{c}}
\toprule
\multirow{2}{*}{\textbf{Crate}} & \multicolumn{6}{c}{\textbf{\# Tests}} & \multicolumn{6}{c}{\textbf{Comp Pass (\%)}} & \multicolumn{4}{c}{\textbf{Exec Pass (\%)}} \\
\cmidrule{2-17}
& RustyUnit & RuMono & RUG & \myadd{SymPrompt} & \myadd{ChatTester} & PALM & RustyUnit & RuMono & RUG & \myadd{SymPrompt} & \myadd{ChatTester} & PALM & RUG & \myadd{SymPrompt} & \myadd{ChatTester} & PALM \\
\midrule
base64        & 50            & 16            & 228          & \myadd{187}  & \myadd{266}  & 629          & \myrep{\tb{100.0}}{100.0} & \myrep{\tb{100.0}}{100.0} & 46.05                     & \myadd{27.81} & \myadd{40.60} & \myrep{48.38}{\tb{48.38}} & 72.38         & \myadd{57.69}      & \myadd{\tb{87.96}} & \myrep{77.78}{\tb{77.78}} \\
bytes         & 47            & 300           & \data\tn{a}  & \myadd{364}  & \myadd{815}  & 1189         & 51.06                     & 57.67                     & \data                     & \myadd{54.40} & \myadd{51.41} & \tb{79.62}                & \data         & \myadd{82.86}      & \myadd{81.25}      & \tb{83.05} \\
fastrand      & 50            & 13            & 195          & \myadd{66}   & \myadd{162}  & 223          & \myrep{\tb{100.0}}{100.0} & \myrep{\tb{100.0}}{100.0} & 55.38                     & \myadd{77.27} & \myadd{54.94} & \myrep{95.96}{\tb{95.96}} & 85.19         & \myadd{86.27}      & \myadd{70.45}      & \tb{90.34} \\
hashbrown     & 50            & \data         & 1391         & \myadd{347}  & \myadd{1310} & 1527         & \myrep{\tb{78.00}}{78.00} & \data                     & 30.70                     & \myadd{6.63}  & \myadd{34.73} & \myrep{41.19}{\tb{41.19}} & \tb{89.23}    & \myadd{69.57}      & \myadd{88.13}      & 82.54      \\
\myadd{http}  & \myadd{\data} & \myadd{213}   & \myadd{1870} & \myadd{673}  & \myadd{1322} & \myadd{2643} & \myadd{\data}             & \myadd{48.83}             & \myadd{1.12}              & \myadd{29.42} & \myadd{41.75} & \myadd{\tb{53.35}}        & \myadd{71.43} & \myadd{88.38}      & \myadd{\tb{91.85}} & \myadd{84.37} \\
indexmap      & 48            & \data         & 1792         & \myadd{335}  & \myadd{1619} & 1876         & \myrep{\tb{100.0}}{100.0} & \data                     & 37.72                     & \myadd{16.72} & \myadd{35.89} & \myrep{50.23}{\tb{50.23}} & \tb{90.38}    & \myadd{85.71}      & \myadd{91.57}      & 86.02      \\
once\_cell    & 42            & 8             & 7            & \myadd{103}  & \myadd{211}  & 299          & 40.48                     & 75.00                     & 28.57                     & \myadd{45.54} & \myadd{43.13} & \tb{89.24}                & 50.00         & \myadd{\tb{91.34}} & \myadd{84.62}      & \myrep{90.94}{\tb{90.94}} \\
\myadd{qsv}   & \myadd{47}    & \myadd{\data} & \myadd{1117} & \myadd{1989} & \myadd{1450} & \myadd{6483} & \myadd{\tb{86.77}}        & \myadd{\data}             & \myadd{45.30}             & \myadd{27.75} & \myadd{52.76} & \myadd{52.43}             & \myadd{67.59} & \myadd{63.77}      & \myadd{\tb{70.07}} & \myadd{65.49} \\
\myadd{quote} & \myadd{50}    & \myadd{57}    & \myadd{205}  & \myadd{31}   & \myadd{87}   & \myadd{113}  & \myadd{\tb{100.0}}        & \myadd{\tb{100.0}}        & \myadd{26.83}             & \myadd{19.35} & \myadd{14.94} & \myadd{67.60}             & \myadd{74.55} & \myadd{\tb{100.0}} & \myadd{84.62}      & \myadd{82.22} \\
rand          & 50            & \data         & 1139         & \myadd{271}  & \myadd{622}  & 840          & \myrep{\tb{76.00}}{76.00} & \data                     & 37.49                     & \myadd{36.90} & \myadd{36.82} & \myrep{47.38}{\tb{47.38}} & \tb{84.31}    & \myadd{61.00}      & \myadd{83.84}      & 82.10      \\
regex         & 50            & 130           & 3732         & \myadd{1004} & \myadd{3359} & 8251         & 94.00                     & \myrep{\tb{96.92}}{96.92} & \myrep{37.73}{\tb{37.73}} & \myadd{27.39} & \myadd{40.19} & 35.48                     & 83.59         & \myadd{72.91}      & \myadd{84.59}      & \tb{85.05} \\
\myadd{ryu}   & \myadd{50}    & \myadd{6}     & \myadd{90}   & \myadd{115}  & \myadd{188}  & \myadd{512}  & \myadd{94.00}             & \myadd{66.67}             & \myadd{\tb{97.78}}        & \myadd{77.39} & \myadd{85.64} & \myadd{87.73}             & \myadd{52.27} & \myadd{59.55}      & \myadd{65.84}      & \myadd{\tb{83.01}} \\
\myadd{serde} & \myadd{44}    & \myadd{\data} & \myadd{5356} & \myadd{954}  & \myadd{1625} & \myadd{2274} & \myadd{\tb{79.55}}        & \myadd{\data}             & \myadd{1.72}              & \myadd{5.97}  & \myadd{19.57} & \myadd{13.32}             & \myadd{82.61} & \myadd{89.47}      & \myadd{89.94}      & \myadd{\tb{93.40}} \\
serde\_json   & 50            & 278           & 2449         & \myadd{955}  & \myadd{2019} & 3116         & 48.00                     & \myrep{\tb{82.09}}{82.09} & 19.44                     & \myadd{15.81} & \myadd{22.44} & \myrep{44.19}{\tb{44.19}} & 81.09         & \myadd{85.74}      & \myadd{85.98}      & \tb{86.09} \\
strsim-rs     & 50            & 9             & 95           & \myadd{85}   & \myadd{140}  & 345          & \myrep{\tb{100.0}}{100.0} & \myrep{\tb{100.0}}{100.0} & 81.05                     & \myadd{60.00} & \myadd{75.00} & \myrep{96.52}{\tb{96.52}} & \tb{92.21}    & \myadd{86.27}      & \myadd{71.43}      & 84.42      \\
\bottomrule
\end{tabular}
\begin{tablenotes}
    \small
    \item[a] The \data\ symbol indicates that the tool was unable to generate tests for these projects.
\end{tablenotes}
\end{threeparttable}
}
\vspace{-0.5cm}
\end{table*}

\textbf{Results.}
Table~\ref{tab:RQ1_a} and~\ref{tab:RQ1_b} present the comparative findings.
\myrep{While classic tools (RustyUnit, RuMono) achieve high compilation rates on simpler tests, PALM emerges as the clear leader among all LLM-based methods. It consistently achieves the highest compilation pass rate in this category, while maintaining a high and stable execution pass rate that is competitive with or superior to other LLM baselines across most projects. For instance, on hashbrown, PALM's compilation rate (41.19\%) is over 6 times that of SymPrompt (6.63\%). In terms of code coverage (Table~\ref{tab:RQ1_b}), PALM's superiority is even more pronounced. It substantially surpasses all other baselines, both classic and LLM-based, across nearly every project. The performance gap is often dramatic; on http, PALM achieves 74.75\% line coverage, more than 40 percentage points higher than the next best LLM-based tool, SymPrompt (29.07\%).}{Analysis failures were noted for RuMono on several crates and for RustyUnit and RUG on one each. Regarding test volume, RustyUnit generated a consistent number of tests (approx. 50), RuMono's output varied by API density, while PALM and RUG produced comparable numbers of shorter tests, with PALM often slightly higher. Classic static approaches (RustyUnit, RuMono) generally exhibited higher compilation rates than LLM-based methods. However, PALM significantly outperformed RUG in compilation success across most crates, with improvements sometimes exceeding 50\%. Excluding RustyUnit (100\% pass rate due to \#[should\_panic]) and RuMono (where pass rate is less directly comparable for fuzzing), PALM generally achieved higher test pass rates than RUG, sometimes by over 10\%. For coverage, LLM-based tools (RUG and PALM) generally surpassed classic methods; PALM demonstrated markedly superior performance, achieving substantially higher line and branch coverage than all other methods across most projects, with increases surpassing 200\%.}
When benchmarked against these manually-authored tests, the original project tests exhibited an average line coverage of \myrep{70.94}{71.30}\% and an average branch coverage of \myrep{66.47}{65.71}\%. In comparison, tests generated by PALM surpassed the line coverage of the original tests in \myadd{almost} half of the evaluated projects, achieving an average of \myrep{72.30}{75.77}\%. Notably, PALM-generated tests demonstrated superior branch coverage in the vast majority of projects, averaging \myrep{69.49}{73.33}\%.

\textbf{Discussion.}
The distinct test generation philosophies influence output characteristics: RustyUnit (SBST) produces comprehensive, longer unit tests, while LLM-based tools generate shorter tests, often targeting single input conditions. RuMono's fuzz target output scales with API density. We observed that, unlike the prolonged execution of RuMono, \myrep{LLM-based tools}{RUG and PALM} typically complete test generation in 2-3 hours, varying with project scale and code complexity. \mydel{While PALM may take slightly longer than RUG for projects with intricate branching, its generation time remains significantly less than typical fuzz testing, highlighting the temporal efficiency of LLMs for this task.}
\myrep{A noteworthy observation is PALM's compilation pass rate, which is lower than that of classic tools like RustyUnit and RuMono. This is a direct consequence of PALM's ambition to generate more complex, human-like tests. Such tests often require intricate object setups and API call sequences, which can trigger the LLM's inherent limitations, such as factual inaccuracies or ``hallucinations,'' making them more prone to compilation errors in a strict language like Rust~\cite{huang2025survey}. This challenge motivates our inclusion of the iterative self-repair mechanism (Section~\ref{sec:gen_repair}). As shown in our ablation study (Table~\ref{tab:RQ2}), this mechanism is highly effective at boosting the final compilation rate and is essential for achieving high test coverage. While the compilation rate is lower than classic tools, PALM generates a significantly larger volume of tests (see Table~\ref{tab:RQ1_a}). This ultimately results in a substantial pool of valid test cases, providing the breadth needed to achieve high coverage.}{Classic tools like RustyUnit and RuMono generally achieve higher compilation success but are not immune to Rust's intricate features (e.g., ownership, types, macros).} Among LLM tools, PALM's superior compilation rates over \myrep{others}{RUG} stem from its use of fine-grained contextual information and iterative self-repair of compilation errors\mydel{, contrasting with RUG's bottom-up input construction}.

Regarding coverage, classic SBST methods can be limited by initial inputs or local optima, and fuzzing often requires extensive runtime. LLMs, with their code understanding, can identify unique execution paths. PALM leverages path-specific requirements to guide generation. While this can sometimes be affected by conflicting constraints or model hallucinations~\cite{huang2025survey}, potentially explaining why improvements in coverage are not always substantial \myadd{on every project}. \myadd{Nevertheless, this approach, particularly its CFG-based analysis and iterative repair mechanism, is the key driver of its overall success. In contrast, the AST-based analysis and test truncation strategy of SymPrompt proved less effective in the complex Rust environment.} PALM's strategy of integrating static analysis for comprehensive function targeting, combined with its higher compilation success rates, allows it to generate tests for a broader set of functions within a project. This synergy of program analysis and effective compilation error correction enables PALM to achieve substantially higher coverage than alternatives, validating its approach and underscoring its potential for practical application.
Across the \myrep{15}{10} evaluated crates, PALM and the human-authored test suites each achieved superior line coverage in \myrep{roughly half of the}{5} instances\mydel{, with comparable line coverage observed for the serde\_json crate}. \myrep{Moreover}{However}, PALM demonstrated higher branch coverage in \myrep{10}{8} out of the \myrep{15}{10} crates, further underscoring its practical utility in software testing workflows. Instances where PALM did not attain superior line coverage were predominantly attributed to the extensive use of macros within those projects, which posed challenges for effective test generation. Conversely, PALM's consistent outperformance in branch coverage is likely due to its\mydel{ inherent} design focus on leveraging branch condition constraints to guide test generation, thereby often achieving more thorough exploration of conditional logic than typical manual efforts.

\begin{table*}[tbp]
\centering
\caption{\myrep{Evaluation of Test Suite Effectiveness via Code Coverage. Bold indicates the best result among all automated test generation tools. Manual coverage is provided as a human-authored reference benchmark.}{Evaluation of different Approaches}}
\label{tab:RQ1_b}
\newcommand{\data}{\textemdash}
\newcommand{\tb}[1]{\textbf{#1}}
\newcommand{\tu}[1]{\uline{#1}}
\newcommand{\tn}[1]{\tnote{#1}}
\resizebox{\linewidth}{!}{
\begin{threeparttable}
\begin{tabular}{l*{7}{c}|*{7}{c}}
\toprule
\multirow{2}{*}{\textbf{Crate}} & \multicolumn{7}{c}{\textbf{Line Cov. (\%)}} & \multicolumn{7}{c}{\textbf{Branch Cov. (\%)}} \\
\cmidrule{2-15}
& RustyUnit & RuMono & RUG & \myadd{SymPrompt} & \myadd{ChatTester} & PALM & Manual & RustyUnit & RuMono & RUG & \myadd{SymPrompt} & \myadd{ChatTester} & PALM & Manual \\
\midrule
base64        & 6.36          & 15.67              & \tb{58.87}    & \myadd{43.33} & \myadd{21.65} & 58.14              & \tu{93.49}         & 10.58         & 17.42              & 58.17         & \myadd{41.18} & \myadd{19.08} & \tb{58.24}         & \tu{87.50} \\
bytes         & 27.53         & 9.01               & \data\tn{a}   & \myadd{56.17} & \myadd{13.40} & \tb{83.93}         & \tu{86.48}         & 28.98         & 10.98              & \data         & \myadd{48.98} & \myadd{10.00} & \tb{75.14}         & 73.31 \\
fastrand      & 58.59         & 35.02              & 60.61         & \myadd{61.28} & \myadd{18.86} & \tb{93.94}         & 63.64              & 62.50         & 7.50               & 66.96         & \myadd{52.63} & \myadd{18.42} & \tb{93.75}         & 49.11 \\
hashbrown     & 7.16          & \data              & 47.50         & \myadd{7.75}  & \myadd{3.31}  & \tb{69.88}         & 53.40              & 5.22          & \data              & 54.21         & \myadd{8.00}  & \myadd{4.86}  & \tb{63.69}         & 60.57 \\
\myadd{http}  & \myadd{\data} & \myadd{33.94}      & \myadd{11.85} & \myadd{29.07} & \myadd{1.77}  & \myadd{\tb{74.75}} & \myadd{64.38}      & \myadd{\data} & \myadd{31.75}      & \myadd{11.31} & \myadd{29.03} & \myadd{2.84}  & \myadd{\tb{73.51}} & \myadd{59.35} \\
indexmap      & 8.23          & \data              & 66.77         & \myadd{18.70} & \myadd{2.63}  & \tb{74.30}         & 56.96              & 8.84          & \data              & 60.86         & \myadd{14.77} & \myadd{2.86}  & \tb{69.47}         & 52.95 \\
once\_cell    & 42.96         & 11.93              & 1.68          & \myadd{50.84} & \myadd{8.11}  & \tb{93.79}         & 80.43              & 49.34         & 11.54              & 1.68          & \myadd{43.82} & \myadd{7.87}  & \tb{92.13}         & 75.00 \\
\myadd{qsv}   & \myadd{9.81}  & \myadd{\data}      & \myadd{28.94} & \myadd{38.43} & \myadd{14.36} & \myadd{\tb{62.18}} & \myadd{\tu{78.83}} & \myadd{8.74}  & \myadd{\data}      & \myadd{28.26} & \myadd{26.65} & \myadd{9.89}  & \myadd{\tb{54.62}} & \myadd{\tu{69.49}} \\
\myadd{quote} & \myadd{3.24}  & \myadd{8.18}       & \myadd{40.91} & \myadd{16.36} & \myadd{4.48}  & \myadd{\tb{77.27}} & \myadd{50.44}      & \myadd{3.24}  & \myadd{8.00}       & \myadd{48.00} & \myadd{22.22} & \myadd{4.41}  & \myadd{\tb{70.34}} & \myadd{57.69} \\
rand          & 14.13         & \data              & \tb{74.29}    & \myadd{42.59} & \myadd{5.76}  & 61.83              & \tu{78.06}         & 14.44         & \data              & \tb{68.06}    & \myadd{36.97} & \myadd{5.37}  & 54.62              & \tu{66.52} \\
regex         & 12.21         & 53.2\myadd{0}      & 46.77         & \myadd{31.75} & \myadd{6.16}  & \tb{55.49}         & \tu{59.64}         & 10.99         & 49.70              & 47.82         & \myadd{32.12} & \myadd{8.72}  & \tb{55.61}         & 54.22 \\
\myadd{ryu}   & \myadd{37.80} & \myadd{\tb{98.27}} & \myadd{68.66} & \myadd{74.65} & \myadd{68.82} & \myadd{93.23}      & \myadd{\tu{100.0}} & \myadd{28.92} & \myadd{\tb{98.04}} & \myadd{61.76} & \myadd{65.57} & \myadd{73.22} & \myadd{90.16}      & \myadd{\tu{98.04}} \\
\myadd{serde} & \myadd{8.11}  & \myadd{\data}      & \myadd{1.35}  & \myadd{14.38} & \myadd{4.90}  & \myadd{\tb{19.46}} & \myadd{\tu{57.00}} & \myadd{9.39}  & \myadd{\data}      & \myadd{1.61}  & \myadd{13.59} & \myadd{6.81}  & \myadd{\tb{20.41}} & \myadd{\tu{55.46}} \\
serde\_json   & 13.88         & 39.45              & 43.36         & \myadd{21.34} & \myadd{5.72}  & \tb{66.65}         & \tu{66.72}         & 16.79         & 40.12              & 50.12         & \myadd{23.18} & \myadd{7.02}  & \tb{73.28}         & 64.01 \\
strsim-rs     & 64.72         & 82.78              & 95.00         & \myadd{80.52} & \myadd{63.61} & \tb{99.72}         & 74.17              & 50.00         & 81.82              & 88.64         & \myadd{71.43} & \myadd{54.55} & \tb{97.40}         & 73.86 \\
\bottomrule
\end{tabular}
\begin{tablenotes}
    \small
    \item[a] The \data\ symbol indicates that the tool was unable to generate tests for these projects.
\end{tablenotes}
\end{threeparttable}
}
\vspace{-0.5cm}
\end{table*}

\subsection{RQ2: Ablation Study}

To optimize test generation, we integrated path constraint-based condition chains, focal context, and LLM-driven error resolution. We conducted an ablation study to evaluate the impact of these components, designing five prompt variants: (i) basic function body and task instructions; (ii) (i) + condition chains; (iii) (i) + context; (iv) (ii) + (iii); (v) (iv) + LLM error fixing, which is \tool. This systematic approach allowed us to quantify the individual contributions of each method.

\textbf{Results.}
Table~\ref{tab:RQ2} details the ablation study results, revealing distinct impacts of different prompting strategies. Basic and context-only prompts generally produced fewer tests than condition chains-only prompts. Context-only prompts typically yielded higher compilation success rates (except on indexmap and strsim-rs) and better test pass rates (except on once\_cell and rand). For code coverage, the general trend was context-only $>$ condition chains-only $>$ basic prompts, though basic prompts occasionally surpassed condition chains-only on some projects. Notably, prompts combining both context and condition chains generally achieved higher coverage than either component in isolation. The efficacy of model self-iteration for repair was evident when comparing configurations with and without this feature, showing universal improvements in compilation rates\textendash exceeding 100\% in some instances\textendash which consequently boosted test pass rates and coverage. Ultimately, the full PALM methodology demonstrated superior performance, achieving the highest compilation rates and crucially, attaining the highest coverage across all evaluated projects.

\textbf{Discussion.}
Our ablation study reveals that basic and context-only prompts, which direct the model to generate an entire test suite in one pass, typically produce fewer tests than condition chain-inclusive prompts designed to generate tests for specific execution paths. Context-only prompts, by incorporating the target method's external dependencies, enable the generation of tests that correctly utilize these dependencies, thus achieving higher compilation success rates. Although condition chain-driven prompts are theoretically poised for higher coverage, their increased propensity for compilation failures often leads to final coverage figures lower than those from context-only prompts. This finding is largely consistent with research on the SymPrompt technique for the GPT-4 model~\cite{ryan2024code}, suggesting that for dialogue-tuned models like GPT-4, contextual information contributes more significantly to effective test generation than explicit condition chains. Nevertheless, combining both context and condition chains generally yields further improvements across most projects. Moreover, leveraging LLMs to repair compilation errors substantially boosts compilation rates on several projects, thereby markedly enhancing the quality of the generated tests.

\subsection{RQ3: Practical Usability}
\label{sec:rq3}

To assess the practical usability of PALM-generated unit tests, we compared their achieved coverage against the existing test coverage in \myrep{15}{10} projects, as detailed in Table\myrep{~\ref{tab:RQ1_b}}{ 4}. We then selected tests that demonstrably increased coverage and submitted them as PRs to these open-source projects.

\textbf{Results.}
The outcomes of the submitted PRs are summarized in the final row of Table~\ref{tab:RQ2}, where `\textbf{P}' denotes PRs still pending developer feedback. In total, 91 tests were submitted across 8 projects. Of these, 80 were accepted and merged, 5 were rejected, and 6 are currently awaiting developer feedback.

\textbf{Discussion.}
The target projects for this evaluation are popular, widely used, and actively maintained, possessing high pre-existing test coverage that can be considered representative of diligent human development efforts. Despite this high baseline, PALM successfully generated tests that increased coverage in 8 of these projects. Among these, we contributed two PALM-generated unit tests to hashbrown, an officially maintained Rust crate that serves as the underlying implementation for HashMap and HashSet in the Rust standard library, thereby enhancing its test coverage and completeness. For projects where PALM did not identify coverage-enhancing tests, our investigation revealed extensive use of macros, a feature for which PALM currently lacks specialized handling, thus limiting its effectiveness in these specific contexts.

Developer interactions during the PR process indicated that while some expressed caution towards LLM-generated code, most were receptive, provided the tests were meaningful and demonstrably increased coverage. The 5 rejected tests were primarily due to developers seeking additional value beyond mere coverage improvement or perceiving the tests as redundant. Several tests were merged after incorporating developer feedback regarding code style, test descriptions, or usage patterns. Feedback is still pending for the remaining 6 tests. Overall, 87.91\% of the submitted tests have been merged by developers. Considering only the tests that received developer review, the merge rate is 94.12\%, underscoring the practical utility and acceptance of PALM-generated tests.

\begin{table*}[tbp]
    \caption{Ablation Result of \tool}
    \label{tab:RQ2}
    \centering
    \newcommand{\data}{\textemdash}
    \newcommand{\tb}[1]{\textbf{#1}}
    \newcommand{\tests}{\# Tests}
    \newcommand{\comp}{Comp Pass (\%)}
    \newcommand{\exec}{Exec Pass (\%)}
    \newcommand{\pline}{Line Cov. (\%)}
    \newcommand{\pbran}{Branch Cov. (\%)}
    \newcommand{\pr}{\# Tests acc/rej}
    \resizebox{\linewidth}{!}{
    \begin{tabular}{ll*{15}{c}}
    \toprule
    \tb{Method} & \tb{Metric} & 
    \tb{base64} & \tb{bytes} & \tb{fastrand} & \tb{hashbrown} & \myadd{\tb{http}} & \tb{indexmap} & \tb{once\_cell} & \myadd{\tb{qsv}} & \myadd{\tb{quote}} & \tb{rand} & \tb{regex} & \myadd{\tb{ryu}} & \myadd{\tb{serde}} & \tb{serde\_json} & \tb{strsim-rs} \\
    \midrule
    \multirow{5}{*}{\centering Basic Prompt}
    & \tests & 212        & 561        & 137        & 1073       & \myadd{1015}       & 1219       & 185        & \myadd{1091}       & \myadd{52}         & 451        & 1199       & \myadd{126}        & \myadd{1056}       & 1438       & 109        \\
    & \comp  & 36.32      & 49.91      & 40.88      & 18.08      & \myadd{36.85}      & 37.90      & 31.35      & \myadd{49.22}      & \myadd{23.08}      & \tb{51.22} & 43.62      & \myadd{69.84}      & \myadd{23.30}      & 19.61      & 72.48      \\
    & \exec  & \tb{96.10} & 83.93      & 83.93      & 77.60      & \myadd{\tb{91.98}} & \tb{95.02} & 89.66      & \myadd{\tb{74.10}} & \myadd{\tb{100.0}} & \tb{85.28} & \tb{89.29} & \myadd{57.95}      & \myadd{\tb{93.90}} & 87.27      & \tb{88.61} \\
    & \pline & 7.59       & 8.84       & 16.84      & 3.42       & \myadd{5.05}       & 1.03       & 3.82       & \myadd{15.90}       & \myadd{12.16}      & 9.97       & 6.30       & \myadd{49.29}      & \myadd{3.50}       & 5.10       & 64.44      \\
    & \pbran & 7.65       & 8.94       & 13.16      & 5.03       & \myadd{5.37}       & 1.36       & 4.49       & \myadd{10.80}       & \myadd{11.86}      & 8.40       & 7.59       & \myadd{49.73}      & \myadd{4.93}       & 6.37       & 58.44      \\
    \cmidrule(lr){2-17}
    \multirow{5}{*}{\centering Condition Chains Only}      
    & \tests & 569        & 1042       & 207        & 1632       & \myadd{2311}       & 1900       & 270        & \myadd{4936}       & \myadd{85}         & 769        & 2643       & \myadd{409}        & \myadd{1873}       & 2978       & 317       \\
    & \comp  & 30.23      & 41.84      & 41.06      & 13.66      & \myadd{28.26}      & 29.53      & 23.33      & \myadd{43.35}      & \myadd{12.94}      & 42.78      & 35.34      & \myadd{73.84}      & \myadd{16.60}      & 15.28      & 82.62     \\
    & \exec  & 70.35      & 78.21      & 76.47      & 81.17      & \myadd{86.68}      & 89.48      & 82.54      & \myadd{63.50}      & \myadd{81.82}      & 79.94      & 79.44      & \myadd{79.47}      & \myadd{90.68}      & 85.62      & 78.24     \\
    & \pline & 31.09      & 18.35      & 15.83      & 2.66       & \myadd{10.93}      & 2.76       & 0.72       & \myadd{25.17}      & \myadd{15.95}      & 11.38      & 7.09       & \myadd{58.58}      & \myadd{3.60}       & 4.81       & 68.61     \\
    & \pbran & 32.94      & 12.57      & 14.06      & 4.18       & \myadd{11.35}      & 4.30       & 1.12       & \myadd{18.49}      & \myadd{16.21}      & 9.66       & 9.03       & \myadd{54.10}      & \myadd{5.42}       & 6.92       & 62.34     \\
    \cmidrule(lr){2-17}
    \multirow{5}{*}{\centering Context Only}
    & \tests & 216        & 647        & 146        & 990        & \myadd{1285}       & 1187       & 209        & \myadd{1274}       & \myadd{69}         & 479        & 1307       & \myadd{160}        & \myadd{1056}       & 1564       & 117        \\
    & \comp  & 47.22      & 68.32      & 85.62      & 17.58      & \myadd{50.12}      & 20.64      & 61.24      & \myadd{\tb{59.03}} & \myadd{23.19}      & 50.94      & 47.59      & \myadd{83.12}      & \myadd{\tb{23.30}} & 31.91      & 71.79      \\
    & \exec  & 84.31      & \tb{83.94} & 77.60      & 81.61      & \myadd{88.82}      & 90.61      & \tb{95.31} & \myadd{73.01}      & \myadd{87.50}      & 68.85      & 81.03      & \myadd{69.92}      & \myadd{\tb{93.90}} & \tb{89.65} & 85.71      \\
    & \pline & 35.74      & 57.01      & 82.83      & 15.71      & \myadd{48.64}      & 22.71      & 62.77      & \myadd{42.92}      & \myadd{40.91}      & 44.46      & 36.47      & \myadd{65.20}      & \myadd{3.50}       & 31.65      & 68.61      \\
    & \pbran & 34.12      & 49.72      & 73.68      & 18.44      & \myadd{47.61}      & 20.90      & 57.30      & \myadd{34.57}      & \myadd{33.33}      & 41.18      & 37.70      & \myadd{65.03}      & \myadd{4.93}       & 37.05      & 64.94      \\
    \cmidrule(lr){2-17}
    \multirow{5}{*}{\centering \tool w/o fix}
    & \tests & 629        & 1189       & 223        & 1527       & \myadd{2643}       & 1876       & 299        & \myadd{6483}       & \myadd{113}        & 840        & 3134       & \myadd{512}        & \myadd{2274}       & 3116       & 345        \\
    & \comp  & 40.38      & 59.71      & 90.58      & 19.19      & \myadd{45.48}      & 25.16      & 56.52      & \myadd{50.96}      & \myadd{25.66}      & 47.02      & 40.62      & \myadd{73.44}      & \myadd{13.28}      & 26.70      & 86.96      \\
    & \exec  & 62.99      & 78.45      & 82.67      & 79.52      & \myadd{84.36}      & 78.60      & 89.94      & \myadd{65.07}      & \myadd{89.66}      & 70.13      & 79.73      & \myadd{76.33}      & \myadd{93.38}      & 85.36      & 81.00      \\
    & \pline & 57.28      & 63.79      & 91.89      & 17.92      & \myadd{60.72}      & 33.01      & 68.02      & \myadd{61.47}      & \myadd{44.45}      & 47.30      & 43.88      & \myadd{74.33}      & \myadd{16.54}      & 31.28      & 89.44      \\
    & \pbran & 58.24      & 58.94      & 92.19      & 22.63      & \myadd{57.95}      & 30.53      & 66.29      & \myadd{54.25}      & \myadd{37.04}      & 45.38      & 43.06      & \myadd{73.77}      & \myadd{15.27}      & 35.86      & 84.42      \\
    \cmidrule(lr){2-17}
    \multirow{5}{*}{\centering \tool}
    & \tests & 629        & 1189       & 223        & 1527       & \myadd{2643}       & 1876       & 299        & \myadd{6483}       & \myadd{113}        & 840        & 3134       & \myadd{512}        & \myadd{2274}       & 3116       & 345        \\
    & \comp  & \tb{48.38} & \tb{79.62} & \tb{95.96} & \tb{41.19} & \myadd{\tb{53.35}} & \tb{50.23} & \tb{89.24} & \myadd{52.43}      & \myadd{\tb{67.60}} & 47.38      & \tb{54.22} & \myadd{\tb{87.73}} & \myadd{13.32}      & \tb{44.19} & \tb{96.52} \\
    & \exec  & 77.78      & 83.05      & \tb{90.34} & \tb{82.54} & \myadd{84.37}      & 86.02      & 90.94      & \myadd{65.49}      & \myadd{82.22}      & 82.10      & 85.05      & \myadd{\tb{83.01}} & \myadd{93.40}      & 86.09      & 84.42      \\
    & \pline & \tb{58.14} & \tb{83.93} & \tb{93.94} & \tb{69.88} & \myadd{\tb{74.75}} & \tb{74.30} & \tb{93.79} & \myadd{\tb{62.18}} & \myadd{\tb{77.27}} & \tb{61.83} & \tb{55.49} & \myadd{\tb{93.23}} & \myadd{\tb{19.46}} & \tb{66.65} & \tb{99.72} \\
    & \pbran & \tb{58.24} & \tb{75.14} & \tb{93.75} & \tb{63.69} & \myadd{\tb{73.51}} & \tb{69.47} & \tb{92.13} & \myadd{\tb{54.62}} & \myadd{\tb{70.34}} & \tb{54.62} & \tb{55.61} & \myadd{\tb{90.16}} & \myadd{\tb{20.41}} & \tb{73.28} & \tb{97.40} \\
    \cmidrule(lr){2-17}
    & \pr    & \tb{P}(1)  & 12/0       & 19/0       & 2/0        & \myadd{\data}      & 39/0       & 3/0   & \myadd{\data} & \myadd{\data} & 5/5     & \data & \myadd{\data} & \myadd{\data} & \data & \tb{P}(5)  \\
    \bottomrule
    \end{tabular}
    }
\vspace{-0.5cm}
\end{table*}

\myadd{\subsection{Qualitative Analysis of Failure Cases}}

\myadd{While PALM demonstrates strong performance, its reliance on an LLM's heuristic understanding means it is not infallible. Failures typically arise when the LLM struggles to synthesize concrete, valid test code from the abstract path constraints provided by our analysis. This synthesis requires inferring a complete test scenario, including object setup and precise API call sequences. We identified two primary failure patterns: compilation failures due to misunderstanding language-specific semantics, and path constraint satisfaction failures due to a lack of causal reasoning.}

\begin{lstlisting}[caption=Compilation failure example., label=code:choice]
pub fn choice<I>(iter: I) -> Option<I::Item>
where I: IntoIterator, I::IntoIter: ExactSizeIterator, {
    with_rng(|r| r.choice(iter))
}
\end{lstlisting}

\myadd{A common compilation failure mode involves the LLM applying programming patterns from other languages that violate Rust's strict ownership rules. For example, when testing the \texttt{choice} function in Code~\ref{code:choice}, which takes an iterator, PALM's LLM component might generate code that moves a Vec into the function and then attempts to reuse it, triggering a ``use of moved value'' error. The LLM fails because its general knowledge, trained heavily on languages like Python or Java where such reuse is common, overrides the specific constraints of Rust's move semantics.}

\lstinputlisting[caption=Path constraint failure example., label=code:reserve]{codes/reserve\_inner.rs}

\myadd{A more subtle failure occurs when a test compiles but does not trigger the intended path because the LLM fails to reason about the side effects of API calls. Consider satisfying the condition \texttt{off >= self.len()} inside the \texttt{reserve\_inner} function in Code~\ref{code:reserve}. Satisfying this condition requires more than just generating input values; it demands a sequence of state-manipulating actions. A correct test must first call a method like \texttt{self.advance()} to modify the object's internal state (\texttt{off}) before calling \texttt{reserve\_inner}. The LLM often fails here because it cannot infer this non-local, causal dependency between an action (\texttt{advance()}) and its future consequence on an internal state check, highlighting a fundamental gap in reasoning about stateful object lifecycles.}

\myadd{\subsection{Cost and Token Analysis}}

\myadd{The practical adoption of LLM-based tools is intrinsically linked to their computational cost. To provide a clear picture of PALM's resource consumption, we analyzed its token usage and associated costs during our evaluation.
On average, generating a single unit test required approximately 1500 input tokens and 300 output tokens. When the iterative repair mechanism was triggered, each repair round consumed an additional 800 tokens on average. For the entire evaluation across all 15 projects, which involved analyzing approximately 40,000 lines of code and generating around 65,000 test cases (including the full ablation study), the total consumption was approximately 800 million tokens. At the prevailing rates for the model used, this translated to a total cost of about \$300. This analysis demonstrates that while not trivial, the cost is manageable for comprehensive academic studies and could be viable for integration into industrial pipelines, especially when applied selectively to critical or complex code modules.}

\myadd{\section{Discussion}}

\myadd{A core design choice in PALM is to use an LLM as a heuristic constraint solver rather than a traditional SMT-based solver used in concolic execution. This choice embodies a fundamental trade-off between formal precision and practical applicability. We do not claim that LLMs are superior to dedicated solvers; for mathematical or logical constraints, the formal guarantees of SMT solvers are unparalleled. PALM, being heuristic, may fail on complex arithmetic constraints, a limitation we acknowledge.
However, PALM's advantages emerge precisely where formal methods often falter: in handling the \textbf{semantic and environmental complexities} of real-world code. For instance, as shown in our motivating example (Section~\ref{sec:example}), PALM successfully generates tests for functions involving concurrency primitives, which are notoriously difficult for concolic engines to handle without specialized, manually-crafted models. More broadly, LLMs excel at satisfying ``semantic'' constraints\textendash such as correct API usage, complex object instantiation, and idiomatic patterns\textendash by leveraging the vast knowledge in their training data. The high compilation success rates and positive developer feedback (Section~\ref{sec:rq3}) from our study underscore this unique capability.}

\myadd{Furthermore, PALM offers crucial advantages in \textbf{scalability and engineering pragmatism}. It directly mitigates the path explosion problem, a key limitation of concolic execution, through its path minimization technique (Section~\ref{sec:path_mini}). By strategically selecting a diverse but minimal set of paths, it avoids the costly endeavor of exhaustive path exploration. This, combined with a significantly \textbf{lower engineering overhead} compared to developing and maintaining a symbolic execution engine, positions PALM as a more pragmatic and readily adaptable alternative. In essence, PALM trades formal precision on narrow constraints for superior handling of semantic complexity, greater scalability, and broader applicability in modern software systems.}

\section{Threats to Validity}

\myrep{\textbf{Threats from Model Selection.}}{\textbf{Model and Language Selection.}}
\myrep{Our study relies exclusively on a single LLM, GPT-4o mini. This presents a threat to external validity, as our findings may not generalize to other models with different architectures or training data. We chose GPT-4o mini for its strong performance and cost-effectiveness, but acknowledge that evaluating PALM with a diverse range of LLMs is a crucial next step. The core techniques of PALM, however, are model-agnostic and designed to be adaptable.}{Our evaluations are conducted on open-source Rust crates using the GPT-4o mini model for test generation. While this may limit PALM's applicability to other LLMs and languages, GPT-4o mini is a state-of-the-art general-purpose model, indicating that PALM's techniques can be applied to different LLMs. Additionally, the methods used in PALM are not specific to Rust and can be adapted to other programming languages.}

\textbf{Subject Programs Selection.}
\myrep{Our subject crates were selected from 15 popular and actively maintained projects on crates.io~\cite{cratesio} and GitHub to cover diverse domains. This selection aims to enhance the applicability of our findings. However, we acknowledge that a study based on 15 projects, despite their diversity and complexity, may not fully represent the entire Rust ecosystem.
Future work should expand this evaluation to an even larger and more varied set of crates}{Our subject crates, carefully selected from 10 crates, demonstrate our commitment to robust research. To ensure the generalizability of our findings, we selected widely-used and actively maintained crates from crates.io~\cite{cratesio} based on download metrics, covering diverse application scenarios. This approach aims to enhance the applicability of our findings to a broader range of projects}.

\myadd{\textbf{Scope of Evaluation.}
Our evaluation primarily focuses on code coverage, a widely-used metric for test suite comprehensiveness. While high coverage is a prerequisite for effective bug finding, a large-scale study on bug detection effectiveness was beyond the scope of this work. We consider this an important direction for future research.}

\section{Related Work}

\textbf{\textit{Search Based Software Testing (SBST).}}
Randoop uses coverage-guided random test generation for Java and includes sanity-checking oracles to detect common bugs like NullPointerExceptions~\cite{pacheco2007feedback, pacheco2007randoop}. EvoSuite is a framework that leverages SBST to generate tests for Java. It supports mutation testing and coverage-guided random test generation~\cite{fraser2011evosuite, fraser2013evosuite, fraser20151600}. CITRUS is a C++ unit testing tool that generates high-coverage test suites by producing random sequences of method calls~\cite{herlim2022citrus}. It is capable of handling challenging technical issues such as template instantiation. Pynguin utilizes SBST to generate regression tests for Python, aiming to ensure code reliability and detect regressions effectively~\cite{lukasczyk2022pynguin}. RustyUnit introduces a search-based approach for automated unit test generation in Rust, utilizing a compiler wrapper and a many-objective genetic algorithm to achieve high code coverage~\cite{tymofyeyev2022search}. PALM conceptually aligns with coverage-driven SBST methods, as both aim to generate test cases guided by coverage criteria. While SBST methods typically employ random search or optimization algorithms to identify suitable test inputs, PALM directs an LLM to generate inputs that satisfy constraints specified for a particular execution path.

\textbf{\textit{Symbolic Execution based Test Generation.}}
KLOVER is the first symbolic execution and automatic test generation tool for C++~\cite{li2011klover, yoshida2017klover}. It is built on top of the symbolic execution engine KLEE~\cite{cadar2008klee} and extends it to support C++. Garg et al. propose a flexible testing framework that alternates between directed-random search and concolic execution to automatically cover hard-to-reach portions of the program~\cite{garg2013feedback}. Chen et al. propose a host-target split concolic testing method for embedded executables, performing symbolic execution on a resource-rich host and concrete execution on the target device, coordinated via cross-debugging, enabling testing without source code in the real environment~\cite{chen2014test}. SyRust leverages a novel logical encoding of Rust’s ownership type system and polymorphic types to synthesize semantically valid test cases, employing symbolic execution and constraint solving to effectively test Rust library APIs while ensuring type safety and proper API call chaining~\cite{takashima2021syrust}. Unlike these approaches, we do not use classic constraint solvers to follow specific paths. Instead, we provide the source code of the focal method to the LLM, delegating the constraint-solving problem to the LLM, which then generates the tests.

\textbf{\textit{LLM based Test Generation.}}
Athenatest fine-tunes pre-trained transformers using paired method-test data and concludes that adding more context in prompts helps generate high-coverage tests~\cite{tufano2020unit}. Mathur et al. utilize T5 and GPT-3 to extract the conversation's context and topic, subsequently generating test cases~\cite{mathur2023automated}. Codamosa employs Pynguin, an SBST tool for Python, and iteratively invokes Codex to generate additional test cases for methods with low coverage~\cite{lemieux2023codamosa}. Yuan et al. perform the first empirical study to evaluate ChatGPT's capability of unit test generation and propose ChatTester, which leverages ChatGPT itself to improve the quality of its generated tests~\cite{yuan2024evaluating}. MuTAP enhances LLM-generated tests through mutation testing and augmented prompts, demonstrating a significant increase in bug detection probability~\cite{dakhel2024effective}. These methods rely on static prompts, whereas PALM combines program analysis \myadd{with LLMs} to generate different prompts for different paths, guiding the model to produce high-coverage test suites. RUG\myadd{~\cite{rug}} introduces a semantic-aware bottom-up approach to leverage LLMs for generating unit tests in Rust, addressing compilation errors through context decomposition and enhancing test coverage by integrating coverage-guided fuzzing, achieving results comparable to human-written tests. SymPrompt~\cite{ryan2024code} is the method most similar to PALM, as it generates corresponding prompts for different paths. However, unlike SymPrompt, we analyze the CFG instead of AST to explore the program's paths better. Meanwhile, instead of merely truncating the model's output, we leverage the model to fix compilation errors in the tests.

\section{Conclusion}

This paper introduces PALM, which combines program analysis with large language models to generate high-coverage unit tests. PALM employs path analysis to obtain the condition chain for each path within the focal method. It then generates a prompt specific to each path by incorporating the context of the function. The model iteratively generates tests and, using provided error information suggests fixes for tests that fail to compile. Our experiments with GPT-4o mini demonstrate that PALM significantly improves both the compilation pass rate and coverage of tests generated by large language models.

\section*{Acknowledgment}

We thank the anonymous reviewers for their helpful feedback and suggestions. This work was supported in part by the National Natural Science Foundation of China under Grant No. 62372225 and No. 62172209, and by the Fundamental Research Funds for the Central Universities (No. 020214380131 and No. ZZKT2025A10).

\bibliographystyle{IEEEtran}
\bibliography{IEEEabrv,ref}

\end{document}